\documentclass[aip,apl,amsmath,amssymb,graphicx,reprint]{revtex4-1}

\usepackage{graphicx}
\usepackage{dcolumn}
\usepackage{bm}

\begin{document}

\title{Characterization of a microwave frequency resonator via a nearby quantum dot }

\author{T. Frey}
\email{freytob@phys.ethz.ch}
\author{P. J. Leek}
\affiliation{Solid State Physics Laboratory, ETH Z\"urich, 8093
Z\"urich, Switzerland}
\author{M. Beck}
\affiliation{Institute for Quantum Electronics, ETH Z\"urich, 8093
Z\"urich, Switzerland}
\author{K. Ensslin}
\author{A. Wallraff}
\author{T. Ihn}
\affiliation{Solid State Physics Laboratory, ETH Z\"urich, 8093
Z\"urich, Switzerland}
\date{\today}

\begin{abstract}
We present measurements of a hybrid system consisting of a microwave transmission-line resonator and a lateral quantum dot defined on a GaAs heterostructure. The two subsystems are separately characterized and their interaction is studied by monitoring the electrical conductance through the quantum dot. The presence of a strong microwave field in the resonator is found to reduce the resonant conductance through the quantum dot, and is attributed to electron heating and modulation of the dot potential. We use this interaction to demonstrate a measurement of the resonator transmission spectrum using the quantum dot.\end{abstract}

\maketitle

The interaction of light and matter is one of the most fundamental processes in physics. One way to explore this area is to use artificial atoms such as quantum dots which offer e.g.~the possibility to tune the energy spacing of the individual electronic states. Using this possibility the resonant absorption of photons by electrons in a quantum dot has been investigated in transport measurements of photon assisted tunneling \cite{osterkamp:98,osterkamp:97}. Cavity quantum electrodynamics (QED), the study of the coupling of matter to light confined in a cavity \cite{haroche}, is traditionally studied with atoms but also with solid state systems such as self-assembled quantum dots \cite{cavitydots1,cavitydots2}. Furthermore the realization of circuit QED \cite{wallraff:04}, in which a single microwave photon is trapped in an on-chip cavity and coherently coupled to a quantum two-level system, has led to significant progress in control and coupling of microwave photons and superconducting qubits. The study of the interaction between the electromagnetic field of such a resonator and a semiconductor quantum dot marks an important step toward realizing a hybrid quantum information processor \cite{losspaper}, in which the advantages of different systems, such as a long relaxation time of the individual qubit \cite{spinpaper} and interaction between distant qubits \cite{majer:07}, could be exploited in one device.

\begin{figure}
	\centering
		\includegraphics[width=1\columnwidth]{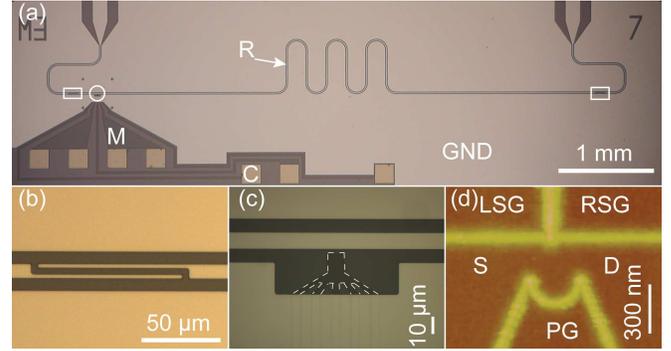}
	\caption{(Color online): (a) Optical micrograph of a microwave resonator (R) with an integrated quantum dot, (GND): ground plane of the resonator, (C): ohmic contact, (M): 2DEG mesa. (b) Magnified view of a coupling capacitor, location on the chip marked with rectangles in (a). (c) Enlarged view of the 2DEG mesa, location on the chip marked with a circle in (a). Edge of the mesa highlighted with a dashed line. (d) AFM picture of the measured quantum dot, realized on the 2DEG mesa.}
	\label{fig:sample}
\end{figure}
The sample, shown in Fig.~\ref{fig:sample} (a), consists of a laterally defined quantum dot positioned at an antinode of the electric field of a microwave transmission-line resonator. The dot is realized on an $\rm{Al_{x}Ga_{1-x}As}$ heterostructure with a two-dimensional electron gas (2DEG) residing at the heterointerface about $35~\rm{nm}$ below the surface. The device is fabricated by three stages of optical lithography followed by local anodic oxidation (LAO) \cite{gustavsson:09} with an atomic force microscope (AFM) to define the quantum dot. In the first of the three lithography steps the mesa for the quantum dot (dark gray parts, labeled M in Fig.~\ref{fig:sample} (a)) is wet etched. Ohmic contacts (labeled C in Fig.~\ref{fig:sample} (a)) are then used to contact the 2DEG. Finally, the microwave resonator and its ground plane (labeled R and GND in Fig.~\ref{fig:sample} (a)) are defined in a lift off process by depositing a bilayer of $3~\rm{nm}$ Ti and $200~\rm{nm}$ Al. The minimum distance from the mesa edge to the resonator center conductor is around $2~\rm{\mu m}$. The coplanar waveguide resonator is designed to have a fundamental frequency $f_0 \approx 7 ~\rm{GHz}$ and is coupled to the input/output lines by two planar finger capacitors (Fig. \ref{fig:sample} (b)). The capacitance, determined in a finite element calculation, corresponds to an external quality factor\cite{goeppl:08} of $Q_{\rm{ext}} \approx 8000$. In Fig.~\ref{fig:sample} (d) the AFM structure is shown which consists of the quantum dot connected by two tunnel barriers to the source (S) and drain (D) contacts used to measure the conductance of the dot. In addition a left side gate (LSG), a plunger gate (PG), and a right side gate (RSG) are used to tune the potential of the quantum dot. The lithographic diameter of the quantum dot is approximately $230~\rm{nm}$. 

The sample is cooled down in a dilution refrigerator with a base temperature of approximately $30~\rm{mK}$, equipped to measure and manipulate the quantum dot at low frequencies (sub-kHz), and to measure the transmission of the resonator in the microwave regime.

\begin{figure}
	\centering
		\includegraphics[width=1\columnwidth]{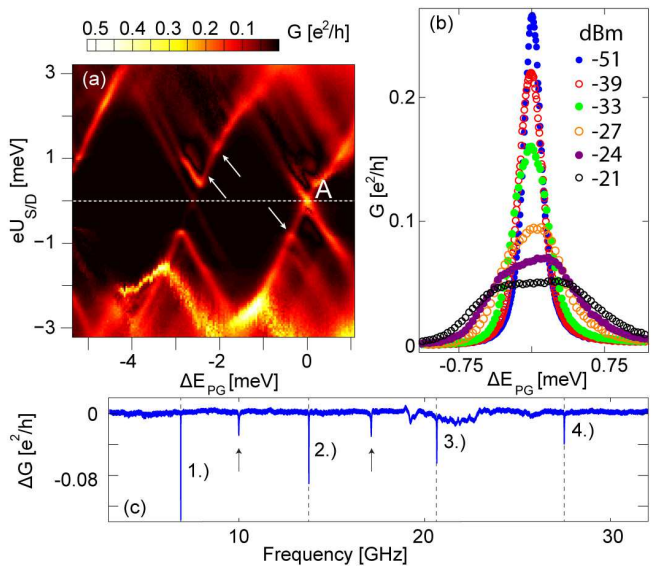}
	\caption{(Color online) (a) Charge stability diagram of the quantum dot without applied microwave power. White arrows indicate excited states. (b) Conductance through the quantum dot as a function of the PG voltage, for different microwave powers applied to the resonator in the vicinity of point A in (a).
	(c) Conductance change $\Delta G$ of the quantum dot as a function of the microwave frequency applied to the resonator.}
	\label{fig:2}
\end{figure}
Coulomb blockade diamonds of the quantum dot with no microwave power applied, measured via lock-in techniques, are shown in Fig.~\ref{fig:2}~(a). The charging energy of the quantum dot is found to be $\Delta E_C \approx 3~\rm{meV}$. The diameter of the dot is estimated with the disc capacitor model $\Delta E_C=e^2/C_{\Sigma}=e^2/4 \epsilon_0 \epsilon_{\rm{GaAs}} d$ to be about $115~\rm{nm}$. Excited state resonances are observable, indicated by white arrows. The typical single-particle level spacing that was resolved is about $350~\rm{\mu eV}$. Using the constant density of states of the 2D system \cite {IhnBuch}, the single-particle level spacing can be used to estimate the diameter of the quantum dot to be about $110~\rm{nm}$. The value is in good agreement with the dot size found using the disc capacitor model. The electron temperature extracted by fitting a thermally broadened Coulomb resonance is $T_e \lesssim 200~\rm{mK}$. Using the maximum of the Coulomb resonance at position A in Fig.~\ref{fig:2} (a), the total tunnel coupling of the quantum dot state to the leads is estimated to be smaller than the thermal energy of the electrons. The measured Coulomb resonances do not have simple thermally broadened line shapes, indicating that the quantum dot is not deep in the single-level transport regime. 

Before studying the interaction between the quantum dot and the microwave field, the transmission of the resonator is characterized using a standard network analyzer. A loaded quality factor \cite{goeppl:08} of $Q_L\approx 2900$ is found for the fundamental mode, at a frequency $f_0 \approx 6.878~\rm{GHz}$. The highest quality factor, that we have reached on bare GaAs with an under-coupled resonator \cite{goeppl:08}, is approximately $10^{4}$.

In a next set of experiments the effect on the conductance of the quantum dot circuit of driving the microwave resonator is investigated. The quantum dot is swept at zero bias through the Coulomb resonance (position A in Fig.~\ref{fig:2} (a)) by changing the voltage on the PG and the dot conductance is recorded via lock-in techniques. The measurement is repeated with the addition of a microwave tone applied to the resonator at $f_0$ for a range of different microwave powers (Fig.~\ref{fig:2} (b)). The applied powers are specified at the output of the microwave generator, and the total damping of the microwave from the generator to the sample is estimated to be about $-30~\rm{dB}$. A reduction in the conductance of the quantum dot and a broadening of the Coulomb resonance with increasing microwave power are observed. We have performed the same measurement on another Coulomb resonance and found similar behavior.

We now investigate the conductance of the quantum dot at the Coulomb resonance (position A, Fig.~\ref{fig:2} (a)), while sweeping the frequency of the signal applied to the resonator. In Fig.~\ref{fig:2} (c) the change of the conductance $\Delta G$ at the Coulomb resonance is plotted as a function of the applied microwave frequency $f$ at a power of $-27~\rm{dBm}$. The change $\Delta G$ is measured relative to the value obtained when the applied microwave frequency is detuned by several GHz from an eigenfrequency of the resonator. Sharp minima in the conductance signal of the quantum dot are observed at frequencies of $6.878~\rm{GHz}$, $13.773~\rm{GHz}$, $20.658~\rm{GHz}$, $27.517~\rm{GHz}$, labeled with 1 to 4 in Fig.~\ref{fig:2} (c). They correspond to the fundamental frequency ($f_0$) and the first three harmonics $(f_{\rm{1}},f_{\rm{2}},f_{\rm{3}})$ of the resonator within an relative error of $E_n=(f_{\rm{n}}-n\cdot f_{0})/f_{\rm{n}} \approx 0.1~\rm{\%}$. The two small additional resonances, marked with arrows, are likely to be caused by sample holder resonances. The conductance measurement through the quantum dot is sensitive enough to observe higher harmonic modes of the microwave resonator that are outside of the frequency range of up to approximately $20~\rm{GHz}$ for which the present microwave setup is designed.
 
\begin{figure}
	\centering
		\includegraphics[width=1\columnwidth]{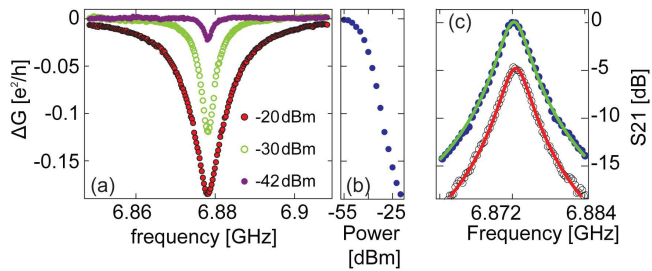}
	\caption{(Color online) (a) Conductance measurement at Coulomb resonance A versus frequency in the vicinity of the fundamental mode, for indicated microwave drive powers. 
	(b) Maximum conductance change $\Delta G$ versus microwave power.
	(c) Resonator spectrum extracted from the dot conductance signal (blue dots) fitted with a Lorentzian line shape (green line), see text. The dark circles are from a network analyzer measurement and are fitted with a Lorentzian line shape as well (red curve). The two curves are offset for clarity.}
	\label{fig:3}
\end{figure}

In a next step the quantum dot conductance is analyzed at a small applied bias of $75~\rm{\mu V}$ in the vicinity of $f_0$. The frequency applied to the feed lines of the resonator is swept and the conductance through the dot is measured with the quantum dot tuned to position A. To relate the influence of the strength of the electromagnetic field to the quantum dot signal, the measurement was repeated for microwave powers ranging from $-55~\rm{dBm}$ to $-20~\rm{dBm}$. In Fig.~\ref{fig:3} (a) the change in conductance $\Delta G$ of the quantum dot is plotted versus the frequency applied to the resonator. The minimum of the conductance is found at the resonance frequency of the resonator and the change of the conductance $\Delta G$ increases with microwave power.
In Fig.~\ref{fig:3} (b) the minima of $\Delta G$, extracted from datasets as shown in Fig.~\ref{fig:3} (a), are plotted as a function of the applied microwave power to establish a relation between the two quantities. The fact that the curve in Fig.~\ref{fig:3} (b) levels off for small microwave power is ascribed to the finite sensitivity of the quantum dot. Note that the smallest input power that is clearly detectable as a change in conductance is $P_{\rm{min}}=-50~\rm{dBm}$, corresponding to a resonator population of $n \approx 15 \cdot 10^4$ photons. Hence a stronger coupling between the two systems would be required to realize a more sensitive detection potentially at the single-photon level.
The conductance scale for the dataset with $P= -20~\rm{dBm}$ in Fig.~\ref{fig:3} (a) is now converted into a power scale using Fig.~\ref{fig:3} (b) and then normalized. A Lorentzian line shape is obtained, blue points in Fig.~\ref{fig:3} (c). The coupling between the dot and the resonator is assumed to be the same over the narrow frequency range covered in the dataset. Also shown with open circles in Fig.~\ref{fig:3} (c) is the $S_{21}$ signal measured with a network analyzer, offset from the other curve by $-5~\rm{dB}$ for clarity. Both curves are fitted with a Lorentzian line shape to extract the loaded quality factor $Q_L$ \cite{goeppl:08}. The obtained values are $Q_L=2896\pm 22$ for the network analyzer measurement, and $Q_L=2890\pm 30$ for the dot conductance based measurement, in very good agreement with each other, supporting the assumption of the constant coupling between resonator and dot. 

We now discuss the potential coupling scenarios between the quantum dot and the microwave resonator. Due to the large extension of the resonator in comparison to the quantum dot there is not only a coupling of the microwave to the quantum dot itself but also to the surrounding 2DEG areas. The direct coupling capacitance between the resonator and the dot in the present geometry is of the order of $1~\rm{aF}$, much smaller than the coupling capacitances between the resonator and the leads, which are in the range of $0.1-1~\rm{fF}$ (estimated from DC characterization measurements and finite element simulations). The conductance peaks in Fig. \ref{fig:2} (c) both decrease in height and broaden with increasing microwave power. These features can be explained as a thermal broadening of the conductance resonance \cite{IhnBuch} which can arise due to heating of the 2DEG by the microwave. In addition the line shapes of the conductance resonances for the higher microwave powers ($-24~\rm{dBm}$ and $-21~\rm{dBm}$, Fig. \ref{fig:2} (c)) indicate that the peak form is a superposition of two resonances. This line shape is consistent with a modulation of the voltage on one or more of the gates and leads of the quantum dot by the microwave. The coupling of the cavity to the quantum dot may therefore be explained as a combination of heating and gate modulation.

In conclusion we have fabricated a quantum dot and a microwave resonator integrated on one chip. Coupling between the two systems could be observed and harmonic modes of the cavity over a frequency range of around $30 ~\rm{GHz}$ could be detected with the quantum dot. Using the quantum dot conductance signal, the quality factor of the fundamental mode of the resonator was extracted.

We thank P.~Studerus, Th.~M\"uller, B.~K\"ung and C.~R\"ossler for technical support. The research was funded by the EU IP SOLID and ETH Zurich.

\end{document}